\documentclass{emulateapj}

\shorttitle{Nature of low $T/|W|$ instabilities}
\shortauthors{Watts, Andersson \& Jones}

\begin{document}

\title{The nature of low $T/|W|$ dynamical instabilities in differentially
rotating stars}

\author{A.L.Watts\altaffilmark{1}, N.Andersson\altaffilmark{2} and
  D.I.Jones\altaffilmark{3}}

\altaffiltext{1}{Laboratory for High Energy Astrophysics, NASA Goddard Space
  Flight Center, Greenbelt, MD 20771, USA; anna@milkyway.gsfc.nasa.gov}
\altaffiltext{2}{School of Mathematics, University of Southampton,
  Southampton SO17 1BJ, UK; na@maths.soton.ac.uk}
\altaffiltext{3}{Center for Gravitational Wave Physics, Pennsylvania State
  University, State College, Pennsylvania 16802-6300, USA}

\begin{abstract}
Recent numerical simulations indicate the presence of dynamical
instabilities of the f-mode in differentially rotating stars 
even at very low values of $T/|W|$, the ratio of
kinetic to potential energy.  In this Letter we argue that these may be
shear instabilities which occur when the degree of differential rotation
exceeds a critical value {\em and} the f-mode 
develops a corotation point associated with the presence of a 
continuous spectrum. Our explanation, which is supported by detailed
studies of a simple shell model, offers a straightforward way of
understanding all of the key features of these instabilities.  
\end{abstract}


\section{Introduction}

Interest in dynamical instabilities of rotating stars is
strong because of the prospects for detecting
gravitational waves from the oscillations of 
neutron stars.  Modelling suggests that for gravitational wave emission to
be at a detectable level we require that the oscillations be
unstable, and hence capable of growth.

One key factor under investigation is the effect
of differential rotation.  Differential rotation may arise in
neutron stars in several circumstances.  
The first is at birth; the latest studies of rotational
core collapse indicate that neutron stars will be born with
strong differential rotation \citep{dm02a, ott04}.  Subsequent accretion of
supernova fallback  
material or material from a companion star \citep{fuj93}, may
also drive differential 
rotation, at least in the surface layers.  Another possibility is that
oscillations may drive the star into differential rotation, via
non-linear effects \citep{rez00,lev01}.  A differentially rotating
massive 
neutron star may also be generated following a binary neutron star merger
\citep{shi00a}. Differential rotation will be relevant provided
that it can be maintained 
in the presence of viscosity or an internal magnetic field, 
which will act to bring the star into uniform
rotation \citep{sha00}.  

Differential rotation has two key effects that are relevant to the
study of dynamical instabilities. Firstly, differentially rotating stars
exhibit a continuous spectrum, with dynamical behaviour that is distinct
from the discrete normal modes of oscillation found in uniformly rotating
stars.  Secondly, differential rotation can lead to the occurrence of
dynamical shear instabilities,  unstable oscillations that do not
exist in uniformly rotating systems \citep{bal85b, luy90}.  

Before proceeding we note some points on terminology.  We state that a mode
is \it{co-rotating} \rm if its pattern speed $\sigma_p = \sigma/m$ is positive, and
\it{counter-rotating} \rm if
its pattern speed is negative (where the 
modes are assumed to behave as $\exp(-i(\sigma t - m\varphi))$).  If
the pattern speed of the mode matches 
the local angular velocity at a point in the star we state that the mode
has a \it{corotation point}\rm. 
The range of frequencies in which modes
possess corotation points is termed the \it{corotation band}\rm.  
All modes with
corotation points are co-rotating, but not all co-rotating modes possess
corotation points.

We will also make reference to the \it{degree of differential
rotation}\rm.  This refers to 
the difference between the maximum and minimum angular velocities
within a differentially rotating star.  Near uniform rotation the
difference is small, and the degree of differential rotation is low. The
difference increases as the degree of differential rotation rises.  
 
\begin{figure}
\centering
\includegraphics[width=8.5cm, height=5cm, clip]{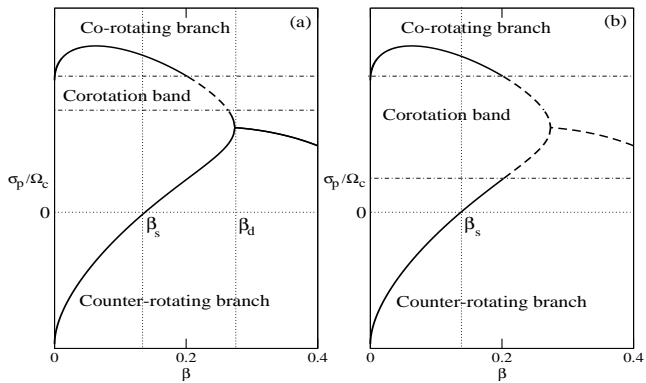}
\caption{Typical dependence of pattern speed $\sigma_p$ on $\beta$ for
  the $l=m=2$ f-modes of a
  differentially rotating stellar model. The quantity
  $\Omega_c$ is discussed in the text. Plot (a) depicts low
  or intermediate differential rotation, plot (b) high
  differential rotation.  The corotation band, traversed
  by the co-rotating f-mode, gets wider as the degree of differential
  rotation rises.}
\label{twfigure1}
\end{figure}

The majority of studies of the oscillations of
differentially rotating systems have focused on the $l=m=2$
f-mode (the so-called bar mode).  As in a uniformly rotating star, the
f-mode is split into co-rotating and
counter-rotating  branches (Fig.\ \ref{twfigure1}).  As the ratio of kinetic
to potential energy, $\beta \equiv T/|W|$, increases, the counter-rotating
mode pattern speed becomes less negative, passing through zero at the point
$\beta = \beta_s$.  It is at this point that secular instability would
arise in the presence of
gravitational radiation reaction.  In the absence of such mechanisms, the
counter-rotating branch pattern speed continues to increase
until it merges with the co-rotating branch at
$\beta = \beta_d$.
This merger gives rise to the well-studied dynamical bar mode instability  
\citep{shi00, new01, yos02, kar03}.   The value of $\beta_d$ is high
($> 0.2$) even for high differential rotation.  

Until recently, the high $\beta$ bar mode instability was the main focus
of attention.  Recent studies by \citet{shi02,shi03} have identified
dynamical bar-mode instabilities in highly differentially rotating
Newtonian polytropes at values of $\beta$ as
low as $\sim 0.01$.  The mechanism
behind these instabilities has yet to be identified.  

The presence of corotation points may be critical.  In differentially
rotating stars the co-rotating branch of the 
f-mode traverses the 
corotation band as $\beta$ is increased (Fig.\ \ref{twfigure1}).  In
this Letter 
we demonstrate that the reported features of the low $\beta$ instabilities
can be explained if
the co-rotating f-mode develops a dynamical shear instability at the
point where it enters the corotation band. This mechanism has been
studied in detail in a simple shell model \citep{wat03,wat04}.
Dynamical instabilities 
associated with the presence of corotation points are also well known
in accretion disc theory \citep{pap85,gol86}.  This would however be the
first occasion on which they have been observed in stellar models.

In section
\ref{cs} we explain the nature of oscillations within the corotation band, and
review the results of a simple model in which the corotation instability mechanism
operates.  In section \ref{main} we show how the mechanism
can account for the unusual features of the low $\beta$ bar-mode instabilities, using the
supporting evidence provided by \citet{shi02, shi03}.

\section{The nature of oscillations in the corotation band}
\label{cs}

In two previous papers \citep{wat03,
wat04} we examined the
nature of oscillations within the corotation band for a
differentially rotating incompressible spherical shell,
by solving the linearized Newtonian
perturbation equations.  We found that within the band there was a
continuous spectrum that is physically distinct from the more familiar
discrete normal modes.  In addition, we found dynamical shear instabilities
that set in when modes developed corotation points.   In this section
we will show that the nature of solutions within the corotation band
remains the same for a more realistic stellar model.

We assume that the star is an inviscid polytrope and that the
perturbations are adiabatic and non-axisymmetric with $\varphi$-dependence 
$\exp(im\varphi)$. We also assume that the perturbations have 
time dependence $\exp(-i\sigma t)$, $\sigma$ being
the frequency.  We define the pressure $P$, density $\rho$, sound
speed $c$,
gravitational potential $\Phi$, and angular velocity $\Omega$.   We
work in the inertial frame and use cylindrical coordinates
($\varpi, \varphi, z$).  Commas are used to denote partial derivatives.
Combining the Euler equations, continuity equation and a polytropic 
equation of state for the perturbations, the Eulerian
perturbation $\delta \phi \equiv \delta P/\rho$ obeys the following
equation \citep{bal85a}
\begin{eqnarray}
\label{pert}
L\phi_{,zz} - \bar{\sigma}^2 \phi_{,\varpi\varpi} +
L\frac{\rho_{,z}}{\rho} \phi_{,z} + \bar{\sigma}^2\phi_{,\varpi}\left[
\frac{L_{,\varpi}}{L} - \frac{\rho_{,\varpi}}{\rho} -
\frac{1}{\varpi}\right] \nonumber {} \\  +  \phi \left[ \bar{\sigma}^2\left(\frac{ L}{c^2} +
\frac{m^2 }{\varpi^2}\right) -
\frac{2m\Omega\bar{\sigma}}{\varpi}\left(\frac{\rho_{,\varpi}}{\rho} +
\frac{\Omega_{,\varpi}}{\Omega} - \frac{L_{,\varpi}}{L}\right)
\right] \nonumber {} \\ -
\frac{\bar{\sigma}^2 L}{c^2} \delta \Phi
=   0 {}\nonumber \\
\end{eqnarray}
where the
corotation parameter $\bar{\sigma} = m\Omega - \sigma$, the 
Lindblad parameter $L = 2\Omega \tilde{\Omega} -
\bar{\sigma}^2$, and the vorticity $\tilde{\Omega} = 2\Omega + \varpi
d\Omega/d\varpi$.  Note that for a barotropic star $d\Omega/dz =
d\Omega/d\varphi = 0$, so that the fluid rotates on cylinders
$\Omega = \Omega(\varpi)$.  

Solution of equation (\ref{pert}) for real $\sigma$ is problematic at the corotation
radius $\varpi = \varpi_c$
where $\bar{\sigma} = 0$, and at the Lindblad radii, $\varpi =
\varpi_L$ where $L=0$. In the neighbourhood of these points the 
character of the solution can be investigated using Frobenius 
analysis. The solution is regular at the Lindblad 
radii. At the corotation radius, although $\phi$ is
continuous its first radial derivative is not.  In general,
there is both a finite step and a logarithmic 
discontinuity in the first derivative at $\varpi_c$ (see \citet{wat03}
for more detail). The additional
degree of freedom introduced by the finite step  allows us to
construct a continuous spectrum of ``eigenfunctions'' for all frequencies
within the corotation band.

The continuous spectrum eigenfunctions found in this more realistic
stellar model have the same mathematical character as those found in
the shell model.  As in this simple model, if one solves the initial
value problem one finds that the collective perturbation associated
with the continuous spectrum is non-singular and
hence physical (\citet{wat04}, see also \citet{cas60,bal84}).

The dynamical shear instabilities found in the simple shell model
occurred where modes crossed into the corotation band, at the
point where modes merged with the continuous spectrum.  Given that the
nature of the continuous spectrum remains the same in the more
realistic model it is legitimate to ask whether the same instability
mechanism might also operate.  In section \ref{main} we will argue
that the low $\beta$ instabilities exhibit all of the key features of
this mechanism.  In order to do this we must first review
some pertinent results from the shell model \citep{wat03, wat04}.

As the degree of differential rotation increases, many of the normal
mode frequencies approach the edge of the expanding corotation band.
Some approach the boundary tangentially and do not cross into
corotation.  Those that cross into corotation at low degrees of differential
rotation remain stable.  Their eigenfunctions generally acquire a
degree of singularity, but can be
distinguished from the rest of the 
continuous spectrum by the fact that
they do not have a finite step in the first derivative.  This led us
to term these solutions ``zero-step solutions''.  We then carried out
numerical time evolutions and an analytical evaluation of the initial
value problem.  When we
produced power spectra of the time evolution data, the zero-step
frequencies stood out just as
clearly as the discrete normal modes outside the corotation band.  The
solutions that one finds when one solves the initial value problem are
in fact non-singular, proving that these 
solutions have physical relevance.

At high differential rotation zero-step solutions are still generated
where modes cross into the corotation band.  In addition, dynamically
unstable (complex frequency) modes can be generated at the crossing
point for certain
rotation profiles.  The eigenfunctions of these dynamically unstable
modes are perfectly regular (recall that our earlier analysis applied
only to real frequencies).  Their pattern speeds are however in
the corotation band. We refer to this instability mechanism as the
corotation instability mechanism.  

\section{Dynamical instabilities}
\label{main}

In reviewing the low $\beta$ instabilities we will focus on the
j-constant rotation law used by Shibata et al.  The j-constant rotation law is 
\begin{equation}
\Omega = \frac{\Omega_c A^2}{A^2 + \varpi^2}
\end{equation}
The parameter $A^2$ sets the degree of differential rotation,
uniform rotation being reached in the limit as $A\rightarrow \infty$.
As $A\rightarrow 0$ the degree of differential rotation increases.
For a given equilibrium model, $\Omega_c$ is a function of the parameter $\beta$.

\begin{figure}
\centering
\includegraphics[width=7cm, clip]{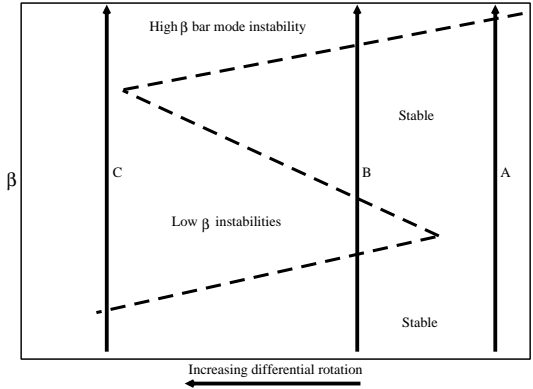}
\caption{A schematic representation of Fig.\ 3 of \citet{shi03}
  showing the regions of parameter space occupied by the high $\beta$
  bar mode and 
  low $\beta$ instabilities.  The lines marked A, B and C are
  discussed in the text.}
\label{twfigure2}
\end{figure}

Following the convention used by Shibata et al and working in dimensionless units where
$0< \varpi < 1$, the corotation band occupies the range of frequencies
\begin{equation}
\frac{mA^2}{A^2 + 1}< \frac{\sigma}{\Omega_c}< m
\end{equation}
Note that the corotation band widens
as degree of differential rotation rises.  
Fig.\ 3 of \citet{shi03},which we
reproduce in schematic form in Fig.\ \ref{twfigure2}, illustrates the
regions of parameter 
space occupied by the high $\beta$ and low $\beta$ instabilities. The
following features  
require explanation.

For low degrees of differential rotation, the low $\beta$
instabilities are not observed.  Dynamical instability does not set in
until high $\beta$,
 where the two branches of the f-mode merge (Track A, Fig.\
 \ref{twfigure2}).  For intermediate degrees of differential rotation,
 the co-rotating
f-mode becomes unstable at low $\beta$.  It stabilises again at intermediate
$\beta$ before merging with the other branch of the f-mode and becoming
dynamically unstable again at high $\beta$ (Track B, Fig.\
\ref{twfigure2}).  For the highest degrees of differential rotation
there is no stable region after the onset of the low $\beta$ 
instability (Track C, Fig.\ \ref{twfigure2}).  With this in mind,
we propose the following explanation of these three feature.

The fact that no
instabilities are observed for stars with low degrees of differential
rotation (Track A, Fig.\ \ref{twfigure2}) suggests that, as for the shell model, this type of shear
instability can only develop if the f-mode crosses into the corotation
band when the degree of differential rotation exceeds a certain
threshold.  For the shell model we were able to determine this threshold
analytically.  Developing a similar criterion for the more realistic model
will however be more difficult. One may, in fact, have to resort
to numerical calculations. Below the threshold the f-mode passes into
the corotation band but remains stable.  Whether the eigenfunction is
then purely regular,
or possesses the singular character of the zero-step solutions that we
found on the shell, is a matter for further study.  

At intermediate degrees of differential rotation (Track B, Fig.\
\ref{twfigure2}) the co-rotating f-mode
enters the corotation band and goes dynamically unstable.  
The lower bound of the low $\beta$ instability region
marks the point at which the
co-rotating f-mode crosses the upper edge of the corotation band.
The upper edge of the low $\beta$ unstable region marks the
point at which the co-rotating f-mode exits the corotation band.  The
system is then stable until the co-rotating branch merges with the
counter-rotating branch at high $\beta$ (see
Fig.\ \ref{twfigure1}(a), 
which illustrates the passage of the co-rotating f-mode through the
corotation band and its exit prior to mode merger).  

As the degree of differential rotation rises (Track C, Fig.\
\ref{twfigure2}), the corotation band gets
wider.  For the scaling used in Fig.\ \ref{twfigure1}, for the
j-constant law, the upper edge of the corotation band remains fixed
whilst the lower edge approaches the zero-axis.  The counter-rotating
branch of the f-mode will 
now enter the corotation band before the co-rotating branch can
emerge (see Fig.\ \ref{twfigure1}(b)).  There is
thus no region of stability.  This raises two
interesting questions.  Does the counter-rotating f-mode also
develop dynamical instability when it enters the corotation band? And
 what becomes of the traditional bar mode instability at high
$\beta$?  If the co-rotating f-mode never stabilises then the traditional mode
merger route is excluded, but we cannot rule out the possibility of
additional instabilities arising via zero-step mergers \citep{bal85b}.

The corotation mechanism provides a simple qualitative explanation of the location
the unstable regions of parameter space.  It also leads to 
quantitative and qualitative predictions that we can test against the
results of the numerical simulations reported by
\citet{shi02,shi03}.  

Firstly, all of the observed low $\beta$ instabilities should have
real parts of frequency that lie within the corotation band. Further
communication with the authors of 
\citet{shi02,shi03} has confirmed that all of the low $\beta$
unstable modes found for the j-constant law lie within the
corotation band (S.Karino, private communication). 

To reproduce the shape of the instability region the co-rotating
f-mode must enter the corotation band at 
lower values $\beta$ and exit the band at higher values of $\beta$ as
the degree of differential rotation rises. Using numerical modelling
we have confirmed that this is the case. 

The third prediction is that the boundaries of the low $\beta$
region of parameter space represent the points at which the
f-mode enters or leaves the corotation band. For models on the lower
edge of the region, we should find that $\sigma_r/\Omega_c =
mA^2/(A^2 + 1)$, while for models on the upper edge of the low
$\beta$ region we should find $\sigma_r/\Omega_c=m$, where
$\sigma_r$ is the real part of frequency.  This
prediction, which would provide the most clearcut confirmation of the
corotation 
mechanism, is however the hardest to test. 

The problem lies in the
growth times.  Although the corotation 
instabilities are genuine dynamical
instabilities (in that they arise in systems for which no dissipative
mechanism is 
present), they only exhibit dynamical (short) growth times deep inside the
corotation band.  The shell model was sufficiently simple that we were able to track
unstable modes with $|\sigma_i/\sigma_r| \sim 10^{-9}$
($\sigma_i$ being the imaginary part of frequency).  It was the
ability to detect these modes with very long
 growth times that allowed us to pinpoint the onset of instability as
 being near 
coincident with the edges of the band. Tracking slow
growing instabilities using the types of numerical codes 
used to investigate realistic stellar models is more difficult.
The code used to generate the results of Shibata et al,
for example, was sensitive to modes with $|\sigma_i/\sigma_r| \sim
10^{-4}$ and larger (S.Karino, private communication).  Such a
limitation on the shell study would have made it impossible for us to
detect the unstable modes until they were well inside the
band. Conclusive tests of the third prediction will therefore require
improved sensitivity to modes with long growth times. Nonetheless
those low $\beta$ instabilities for which growth times have been 
published do exhibit the expected behavior in that
the growth times are shortest in the center of the band, getting
longer towards the edges (see for example Fig.\ 4 of
\citet{shi02}).

Although there are other instability mechanisms associated with the corotation band
\citep{bal85b, luy90, pap85}, all have limitations as an explanation for
the low $\beta$ bar mode instabilities. In these scenarios the co-rotating f-mode would enter the corotation band as 
either a zero-step or decaying mode (although no decay is observed),
merging inside the band 
with another zero-step or decaying mode to go unstable.  What it could be
merging with is however unclear:  the counter-rotating f-mode is outside the band at low
$\beta$, and the r-modes never enter \citep{kar01}. In addition it is
not clear whether the unstable mode could retain a bar-like character if
it arose through merger with another mode type; the corotation
mechanism, by contrast, permits bar-like instability as the f-mode
would not merge with any other mode.  

The corotation mechanism may also be responsible for the low $\beta$ $m=1$ spiral
instabilities observed by several authors, since the unstable modes of
\citet{pic96} and \citet{cen01} are all within the corotation band.
This is a topic for future work.

\section{Conclusions}
The presence of shear instabilities in differentially rotating stars
is conceptually very interesting. It is also possible that these instabilities
will turn out to be of considerable astrophysical significance. These
instabilities may limit the range of astrophysical rotation laws. 
In addition the dynamically unstable modes may lead to 
interesting gravitational wave signals. It is 
worth noting that, while it is difficult to 
envisage scenarios that lead to the large value of $\beta$ 
required for the more familiar bar-mode instability, one can easily
imagine situations where the low $\beta$ shear instabilities 
will operate. 

We have suggested a scenario, motivated by our analysis of the
differentially rotating shell, that explains the low $\beta$ f-mode
instabilities observed by Shibata and collaborators.  In this scenario dynamical
shear instabilities arise when the co-rotating f-mode enters the corotation
band and the degree of differential rotation exceeds a certain threshold
value.  The corotation mechanism provides a straightforward
explanation for all of the unusual features of the instabilities
observed in numerical simulations, and is supported by all available
data.  Further tests of the mechanism will require high precision
studies on the onset of the instability, a task that will require
codes sensitive to dynamically unstable modes with very long growth
times.  

\acknowledgments
ALW is a NRC RRA fellow at NASA
GSFC.  NA acknowledges
support from the Leverhulme Trust in the form of a prize fellowship. DIJ is supported by the Center for Gravitational Wave Physics
under NSF cooperative agreement PHY 01-14375. We thank S.Karino for
helpful comments.

\clearpage


\begin{thebibliography}{99}
\bibitem[\protect\citeauthoryear{Balbinski}{1984}]{bal84}
Balbinski E., 1984, \mnras, 209, 145
\bibitem[\protect\citeauthoryear{Balbinski}{1985a}]{bal85a}
Balbinski E., 1985, \aap, 149, 487
\bibitem[\protect\citeauthoryear{Balbinski}{1985b}]{bal85b}
Balbinski E., 1985, \mnras, 216, 897
\bibitem[\protect\citeauthoryear{Case}{1960}]{cas60}
Case, K.M., 1960, Physics of Fluids, 3, 143
\bibitem[\protect\citeauthoryear{Centrella et al}{2001}]{cen01}
Centrella J.M., New K.C.B., Lowe L.L., Brown D.J., 2001, \apj, 550, L193 
\bibitem[\protect\citeauthoryear{Dimmelmeier, Font \&
M\"uller}{2002}]{dm02a}
Dimmelmeier H., Font J.A., M\"uller E., 2002,\aap, 388, 917
\bibitem[\protect\citeauthoryear{Fujimoto}{1993}]{fuj93}
Fujimoto M.Y., 1993, \apj, 419, 768
\bibitem[\protect\citeauthoryear{Goldreich, Goodman \&
Narayan}{1986}]{gol86}
Goldreich P., Goodman J., Narayan R., 1987, \mnras, 221, 339
\bibitem[\protect\citeauthoryear{Karino \& Eriguchi}{2003}]{kar03}
Karino S., Eriguchi Y., 2003, \apj, 592, 1119
\bibitem[\protect\citeauthoryear{Karino, Yoshida \& Eriguchi}{2001}]{kar01}
Karino S., Yoshida S., Eriguchi Y., 2001, \prd, 64, 024003
\bibitem[\protect\citeauthoryear{Levin \& Ushomirsky}{2001}]{lev01}
Levin Y., Ushomirsky G., 2001, \mnras, 322, 515
\bibitem[\protect\citeauthoryear{Luyten}{1990a}]{luy90}
Luyten P.J., 1990, \mnras, 242, 447
\bibitem[\protect\citeauthoryear{New \& Shapiro}{2001}]{new01}
New K.C.B., Shapiro S.L., 2001, \apj, 548, 439
\bibitem[\protect\citeauthoryear{Ott et al}{2004}]{ott04}
Ott C.D., Burrows A., Livne E., Walder R., 2004, \apj, 600, 834
\bibitem[\protect\citeauthoryear{Papaloizou \& Pringle}{1985}]{pap85}
Papaloizou J.C.B., Pringle J.E., 1985, \mnras, 213, 799
\bibitem[\protect\citeauthoryear{Pickett, Durisen \& Davis}{1996}]{pic96}
Pickett B.K., Durisen R.H., Davis G.A., 1996, \apj, 458, 714
\bibitem[\protect\citeauthoryear{Rezzolla et al}{2000}]{rez00} 
Rezzolla L., Lamb F.L., Shapiro S.L., 2000, \apj, 531, L139
\bibitem[\protect\citeauthoryear{Shapiro}{2000}]{sha00}
Shapiro S.L., 2000, ApJ, 544, 397
\bibitem[\protect\citeauthoryear{Shibata \& Uryu}{2000}]{shi00a}
Shibata M., Uryu K., 2000, \prd, 61, 064001
\bibitem[\protect\citeauthoryear{Shibata, Baumgarte \&
Shapiro}{2000}]{shi00}
Shibata M., Baumgarte T.W., Shapiro S.L., 2000, \apj, 542, 453
\bibitem[\protect\citeauthoryear{Shibata, Karino \& Eriguchi}{2002}]{shi02}
Shibata M., Karino S., Eriguchi Y., 2002, \mnras, 334, L27
\bibitem[\protect\citeauthoryear{Shibata, Karino \& Eriguchi}{2003}]{shi03}
Shibata M., Karino S., Eriguchi Y., 2003, \mnras, 343, 619
\bibitem[\protect\citeauthoryear{Watts et al}{2003}]{wat03}
Watts A.L., Andersson N., Beyer H., Schutz B.F., 2003, \mnras, 342,
1156
\bibitem[\protect\citeauthoryear{Watts, Andersson \& Williams}{2004}]{wat04}
Watts A.L., Andersson N., Williams R.L., 2004, \mnras, 350, 927
\bibitem[\protect\citeauthoryear{Yoshida et al}{2002}]{yos02}
Yoshida S., Rezzolla L., Karino S., Eriguchi Y., 2002, \apj, 568, L41

\end{thebibliography}
\end{document}